\documentclass[preprint,showpacs,preprintnumbers,amsmath,amssymb]{revtex4}
\usepackage{graphicx}
\usepackage{dcolumn}
\usepackage{bm}
\usepackage{mathrsfs}
\usepackage{amsmath}

\begin{document}
\title{Corrections to the Nonrelativistic Ground Energy of a Helium Atom
\footnote{Supported by the National Natural Science Foundation of China, grant No. 10275030.}}

\author{Yi-Shi Duan}
\author{Yu-Xiao Liu}
\thanks{Corresponding author}\email{liuyx01@st.lzu.edu.cn}
\author{Li-Jie Zhang}
\affiliation{Institute of Theoretical Physics, Lanzhou University, Lanzhou 730000, P. R. China}

\begin{abstract}
Considering the nuclear motion, the authors give out the nonrelativistic ground energy of a helium atom by using a simple but effective variational wave function with a flexible parameter $k$. Based on this result, the relativistic and radiative corrections to the nonrelativistic Hamiltonian are discussed. The high precision value of the helium ground energy is evaluated to be $-2.90338$ a.u., and the relative error is 0.00034\%.
\end{abstract}


\maketitle

As is well known, studying the typical Coulomb three-body bound-state problem such as a helium atom is a very basic and important problem in the field of atomic physics ever since. Much work has been carried out by some approximate methods. One of the earliest variational calculations has been performed by Hylleraas \cite{Hylleraas1929} in 1929. Since this time, many physicists have contributed incremental steps in this endeavor, often trying to use mathematical insight for advantage \cite{Fock1954}-\cite{Korobov2001}. The strongest line of theory has been focused on the analytic properties of the wave function, especially following the work of Fock in 1954 \cite{Fock1954}, which showed a weak logarithmic singularity at the three-particle coalescence. The recent works of Korobov and Charles Schwartz ect get very high precision result for the nonrelativistic ionization \cite{Korobov2000,Korobov2002,Schwartz2002,GUAN Xiao-Xu}.

But most of these authors mainly lay their works on the precision and convergence of the variational wave function except for the correction to Hamiltonian of a helium atom. Even the motion of nucleus is not considered, so the variational results are lower than the experimental value \cite{Fock1954}-\cite{ChenYH2003}. In this paper, we first construct a simple but effective variational wave function with a flexible parameter k. Then, considering the motion of nucleus and the relativistic and radiative effects, we give out all the corrections to the nonrelativistic ground energy. Finally, the ground energy of a helium atom is worked out with the variational method and perturbation method.

It is well known that the Schr\"{o}dinger ground energy $E$ and the corresponding wave function $U$ are found as a solution of the variational problem
\begin{equation}\label{E}
E = \min_{U} \frac{\int {U^\ast H U d\tau}}%
                  {\int {U^\ast U d\tau}}.
\end{equation}
For a helium atom, we use a simple but effective variational wave function, which contains only even powers of $t$, as follows
\begin{equation}\label{U}
U(\textbf{r}_1,\textbf{r}_2)=\varphi(ks,kt,ku) %
=e^{-ks}\sum_{lmn} C_{lmn}(ks)^{l}(kt)^{2m}(ku)^{n},
\end{equation}
\begin{equation}\label{stu}
s=r_{1}+r_{2},~%
t=-r_{1}+r_{2},~%
u=r_{12},
\end{equation}
where $r_1$ and $r_2$ denote the positions of the electrons with respect to the nucleus; $s$, $t$ and $u$ are called Hylleraas coordinates \cite{Hylleraas1929}; $k$ is a flexible scaling parameter. The function may be expected to converge rapidly for medium-sized of $s$, $t$ and $u$. The Hamiltonian is taken in following nonrelativistic approximation in many papers \cite{Fock1954}-\cite{ChenYH2003}
\begin{equation}\label{Hinf}
H_{\infty} = \frac{\textbf{p}_{1}^{2}}{2} %
            +\frac{\textbf{p}_{2}^{2}}{2} %
            -\frac{Z}{r_1} -\frac{Z}{r_2} %
            +\frac{1}{r_{12}},
\end{equation}
where $\textbf{p}_{1,2}$ are the momenta of the electrons and $Z=2$ is the nucleus charge (in units of the proton charge). And the variational results in References \cite{Fock1954}-\cite{ChenYH2003} are lower than the experimental value, which shows that the Hamiltonian form in Eq. (\ref{Hinf}) is not accurate enough. So the movement of the nucleus should not be ignored, and the nonrelativistic Hamiltonian can be represented as follows
\begin{equation}\label{H0}
H_{0} =      \frac{\textbf{p}_{1}^{2}}{2} %
            +\frac{\textbf{p}_{2}^{2}}{2} %
            +\frac{\textbf{P}^{2}}{2M} %
            -\frac{Z}{r_1} -\frac{Z}{r_2} %
            +\frac{1}{r_{12}}.
\end{equation}
Here $\textbf{P}=-\textbf{p}_1-\textbf{p}_2$ is the momentum of the nucleus and the finite nucleus-to-electron mass ratio $M\equiv m_\alpha/m_e= 7294.299508(16)$. From the later computational results, it can be seen clearly that the variational energy is higher than the experimental value.

To get more accurate result, we take into account the relativistic and radiative effects. The following operator describes the Breit $\alpha^2$ correction to the nonrelativistic Hamiltonian \cite{Bethe1957}
\begin{equation}
\label{H^2}
H^{(2)} = H_1 + H_2 + H_3 + H_4 + H_5,
\end{equation}
\begin{equation}
\label{H_1}
H_1 = -\frac{\alpha^2}{8}(\textbf{p}_1^4 + \textbf{p}_2^4),
\end{equation}
\begin{equation}
\label{H_2}
H_2 = - \frac{\alpha^2}{2}%
\left[
 \frac{\textbf{p}_1 \cdot \textbf{p}_2}{r_{12}}
 +\frac{(\textbf{r}_{12}\cdot\textbf{p}_1)
 (\textbf{r}_{12}\cdot\textbf{p}_2)}{r_{12}^3}
\right],
\end{equation}
\begin{equation}
\label{H_3}
H_3 = \frac{\alpha ^2}{2}\left\{
      \left[
        \textbf{L}_1 \times \textbf{p}_1
        +2\frac{\textbf{r}_{12} \times \textbf{p}_2}{r_{12}^3}
        \right] \cdot\textbf{S}_1
      +\left[\textbf{L}_2 \times\textbf{p}_2
        +2\frac{\textbf{r}_{21} \times\textbf{p}_1}{r_{12}^3 }
        \right] \cdot\textbf{S}_2 \right\},
\end{equation}
\begin{equation}
\label{H_4}
H_4 = \frac{i\alpha^2}{4}(\textbf{p}_1 \cdot\textbf{L}_1
      +\textbf{p}_2 \cdot\textbf{L}_2 ),
\end{equation}
\begin{equation}
\label{H_5}
H_5 = \alpha^2 \left\{-\frac{8\pi}{3} (\textbf{S}_1\cdot\textbf{S}_2)\delta(\textbf{r}_{12})
+\frac{1}{r_{12}^3}
\left[\textbf{S}_1\cdot\textbf{S}_2
-\frac{3(\textbf{S}_1\cdot\textbf{r}_{12}) (\textbf{S}_2\cdot\textbf{r}_{12})}{r_{12}^2}\right]' \right\}.
\end{equation}
Here $\textbf{L}_1 = -\nabla_{1} V ~(V=Z/r_1 + Z/r_2 - 1/r_{12})$ is the Coulomb field due to nucleus plus the second electron. The physical significance of the various terms in Eq. (\ref{H^2}) is as follows:

$H_1$ is the relativistic correction due to the ``variation of mass with velocity" (which does not depend on electron spin). $H_2$ corresponds to the classical relativistic correction to the interaction between the electrons. This correction is due to the retardation of the electro-magnetic field produced by an electron. $H_3$ is the interaction between the spin magnetic moment and the orbital magnetic moment of the electrons (spin-orbit coupling). $H_4$ is a term characteristic of the Dirac theory, which is also present in the Hamiltonian for a single electron in an electric field. $H_5$ represents the interaction between the spin magnetic dipole moments of the two electrons.

For the ground state of helium-like atoms, $l$, $s$ and $j$ are all zero and there is no fine structure splitting. Nevertheless, the operators $H_1$ to $H_5$ in Eq. (\ref{H^2}) contribute relativistic corrections to the nonrelativistic energy eigenvalue of relative order $(Z\alpha)^2$ and $Z\alpha^2$. These corrections are only the leading terms in an expansion in powers of $\alpha$ and $Z\alpha$. And we explicitly take into consideration that the spin of the nucleus and the total spin of electrons are both equal to zero. In particular, we replace the product of the electron spin operators $\textbf{S}_{1}\cdot \textbf{S}_{2}$ by its eigenvalue in the single state, $-3/4$. The expectation values of the operators $H_1$ to $H_5$ for the ground state of a helium atom with nuclear charge $Z$ are
\begin{equation}
\label{E_1}
E_1=-\frac{\alpha^2}{8}(<\textbf{p}_1^4>+<\textbf{p}_2^4>),
\end{equation}
\begin{equation}
\label{E_2}
E_2 = - \frac{\alpha ^2}{2}\left[<
\frac{\textbf{p}_1\cdot\textbf{p}_2}{r_{12}}>
+<\frac{(\textbf{r}_{12}\cdot\textbf{p}_1)(\textbf{r}_{12}
\cdot\textbf{p}_2 )}{r_{_{12} }^3 } > \right],
\end{equation}
\begin{equation}
\label{E_3}
E_3 = 0,
\end{equation}
\begin{equation}
\label{E_4}
E_4 = \pi \alpha ^2 < Z\frac{\delta(\textbf{r}_1 )
      + \delta(\textbf{r}_2)}{2}- \delta(\textbf{r}_{12})> ,
\end{equation}
\begin{equation}
\label{E_5}
E_5 = 2\pi \alpha ^2 < \delta(\textbf{r}_{12})>.
\end{equation}
The angle brackets in Eqs. (\ref{E_1}) to (\ref{E_5}) and below denote the average value over the nonrelativistic ground state variational wave function. It can be shown that the expectation value $E_{2}$ of the operator $H_{2}$ vanishes if any wave function of product form $U = u(r_{1})u(r_{2})$ is used, i.e. both for the hydrogen-like and for the Hartree wave functions. If a more accurate function, which includes the effects of polarization, is used, then a finite (but numerically small) value is obtained for $E_{2}$. To simplify the presentation, we take $E_{2}=0.$ So, the total Breit $\alpha ^{2}$ correction to the nonrelativistic value is
\begin{equation}
\label{deltaE^2}
\delta E^{(2)} = \alpha^2 <-\frac{\textbf{p}_1^4
        +\textbf{p}_2^4}{8}+ \pi Z\frac{\delta(\textbf{r}_1)
        +\delta(\textbf{r}_2)}{2}
        +\pi\delta(\textbf{r}_{12})>.
\end{equation}

Order $\alpha ^{3}$ correction to the nonrelativistic energy can be
represented as follows \cite{Yelkhovsky2001}
\begin{eqnarray}\label{deltaE^3}
\delta E^{(3)} &=& \alpha^3\left\{
 {\frac{4Z}{3}\left(-2\ln\alpha-\beta+\frac{19}{30}\right)
 <\delta(\textbf{r}_1)+\delta(\textbf{r}_2)>}
\right.\nonumber \\
&&\left.
+\left(\frac{14}{3}\ln\alpha+\frac{164}{15}\right) \delta(\textbf{r}_{12})
 +\frac{7}{3\pi}<\frac{\ln r_{12}+\gamma}{r_{12}^2}%
  i\textbf{n}\cdot\textbf{p}>
\right\}.
\end{eqnarray}
Here
\begin{equation}
\textbf{n} = \frac{\textbf{r}_{12}}{r_{12}},~
\textbf{p} = -i\nabla = - i\frac{\partial }{\partial \textbf{r}_{12} },
\end{equation}
$\gamma=0.5772$ is the Euler constant and $\beta=4.3700392$ is the helium Bethe logarithm \cite{Kabir1957}.

Now, we give out the calculations and results of the helium ground energy.

Firstly, we shall discuss the ground state expectation values of the Dirac delta-function, the square of the kinetic energy operator and the nonrelativistic energy, which evaluated by using various wave functions.

The expectations of the three-dimensional Dirac delta-function $\delta (\textbf{r}_{1})$ and $\delta(\textbf{r}_{12})$ are
\begin{equation}
\label{delta r1}
 <\delta(\textbf{r}_1)>
 =\int{d\tau_2 U^2(0,\textbf{r}_2)}
 =4\pi\int\limits_0^\infty {r^2\varphi ^2(r,r,r)} dr,
\end{equation}
\begin{equation}
\label{delta r12}
 <\delta (\textbf{r}_{12})>
 = \int{d\tau_1}U^2(\textbf{r}_1,\textbf{r}_1)
 = 4\pi\int\limits_0^\infty {r^2\varphi ^2(2r,0,0)} dr.
\end{equation}
It is very convenient to evaluate them in spherical coordinates.

For the square of the kinetic energy operator for electron 1, or
$\textbf{p}_{1}^{4}=\nabla_{1}^{4} \equiv \Delta_{1}^{2}$, we first note the following equation
\begin{equation}
\label{eq_U}
\nabla_1 \cdot \left[
U \nabla_1 (\Delta _1 U) - (\Delta _1 U)\nabla_1 U \right]
= U(\Delta _1^2 U) -
(\Delta _1 U)^2.
\end{equation}
For any analytic function $U$ that falls off exponentially at large distances, the integral of the left side of Eq. (\ref{eq_U}) over the whole $\textbf{r}_{1}$-space must vanish (from Gauss' theorem). Then there are two alternative forms for the expectation value of $\textbf{p}_{1}^{4}$
\begin{equation}
\label{eq_p14}
 <\textbf{p} _1^4 >
 = \int {d\tau _1 d\tau _2 U\Delta _1^2 U}
 = \int {d\tau _1 d\tau _2 (\Delta _1 U)^2}.
\end{equation}
Great care must be taken if the first form in Eq. (\ref{eq_p14}) is used. For the exact wave function, $\Delta_{1}U$ behaves like $Z/r_{1}$ or like $1/r_{12}$ if $r_{1}$ or $r_{12 }$approaches to zero, just like the potential energy in the total Hamiltonian. $\Delta_{1}^{2}U$ then has a delta-function type of singularity at $r_{1}=0$ and $r_{12}=0$ and a wrong answer would be obtained if the first integral in Eq. (\ref{eq_p14}) were evaluated by a limiting process which excludes an infinitesimal region around the origin and around $r_{12}=0$. The second form of Eq. (\ref{eq_p14}) is free from these difficulties and is, in any case, easier to evaluate in practice. Considering the exchange symmetry of the wave function, we have $<\delta (\textbf{r}_{1})> = <\delta(\textbf{r}_{2})>$ and $<\textbf{p}_{1}^{4}>=<\textbf{p}_{2}^{4}>$.

The expression of the nonrelativistic ground energy has the following form
\begin{equation}
\label{E_0}
E_0 = < H_0 >  = \frac{K}{W}k^2 + \frac{P}{W}k,
\end{equation}
where $K$ is kinetic energy, $P$ is potential energy and $W$ is a normalization factor. The function $E_{0}(k)$ is a quadratic parabola that take the minimum at $k=-P/2K$. When the number of basis functions $N$ is large enough, the value of $k$ will fluctuate between 2.0451 and 2.0452. For example, we get $k=2.0451486913735$ when $N=50$.

Secondly, we will present the computational values for all the contributions to the helium ground energy in atomic units ($m_{p}$=1, $\hbar$=1 and $e$=1). The particular numerical results and their comparisons with the experimental value are collected in Table 1 and Table 2. $E_{\infty }$ and $E_{0}$ are calculated by variational method, $\delta E_{chr}$, $\delta E^{(2)}$ and $\delta E^{(3)}$ by perturbation method.

\vskip 8mm
\textbf{Table 1.} Nonrelativistic energies $E_{\infty }$ and$ E_{0}$ for the helium ground state and their comparisons with the experimental value. $E_{\infty}$ corresponds to the expectation value of $H_{\infty}$, $E_{0}$ to the expectation value of $H_{0}$ and $E_{exp}$ to the experimental value, which equals to $-2.90338629$ \cite{Radzing1985} (in atomic units). $\Delta E_{\infty} =E_{\infty} - E_{exp}$ and $\Delta  E_{0} =E_{0} - E_{exp}.$
\begin{center}
\begin{tabular}{|c|c|c|c|c|}
  \hline\hline
  $N$ & $E_{\infty}$ & $\Delta E_{\infty }$ & $E_{0}$ & $\Delta E_{0}$ \\
  \hline
  ~~~20 ~~~&~~~ -2.90370938 ~~~&~~~  -0.00032309 ~~~&~~~ -2.90328962 ~~~&~~~ 0.00009667~~~ \\
  ~~~30 ~~~&~~~ -2.90371945 ~~~&~~~  -0.00033316 ~~~&~~~ -2.90329969 ~~~&~~~ 0.00008660~~~ \\
  ~~~40 ~~~&~~~ -2.90372103 ~~~&~~~  -0.00033474 ~~~&~~~ -2.90330128 ~~~&~~~ 0.00008501~~~ \\
  ~~~50 ~~~&~~~ -2.90372124 ~~~&~~~  -0.00033495 ~~~&~~~ -2.90330389 ~~~&~~~ 0.00008240~~~ \\
  \hline
 \end{tabular}
\end{center}

\vskip 8mm

\textbf{Table 2.} The relativistic and radiative corrections to the helium nonrelativistic ground energy and the helium ground energy $E$. $E= E_{0}+\delta E^{(2)}+\delta E^{(3)}, \Delta E=E-E_{exp}.$
\begin{center}
\begin{tabular}{|c|c|c|c|c|}
  \hline\hline
$N$ & $\delta E^{(2)}$ & $\delta E^{(3)}$ & $E$  & $\Delta E$\\
\hline
~~~20 ~~~&~~~ -0.00009584 ~~~&~~~0.00002238 ~~~&~~~-2.90336 ~~~&~~~ 0.00003~~~ \\
~~~30 ~~~&~~~ -0.00009610 ~~~&~~~0.00002239 ~~~&~~~-2.90337 ~~~&~~~ 0.00002~~~ \\
~~~40 ~~~&~~~ -0.00009586 ~~~&~~~0.00002238 ~~~&~~~-2.90337 ~~~&~~~ 0.00002~~~ \\
~~~50 ~~~&~~~ -0.00009586 ~~~&~~~0.00002238 ~~~&~~~-2.90338 ~~~&~~~ 0.00001~~~ \\
\hline
\end{tabular}
\end{center}

\vskip 8mm

It can be seen from Table 1 that the variational ground energy $E_{\infty }$ is lower than the experimental value and the error will become larger increasing with the number of basis functions$ N$, which indicates that the Hamiltonian in Eq. (\ref{Hinf}) is not accurate enough. However, the case of $E_{0}$ is opposite to that of $E_{\infty }$ and it approaches to the experimental value when $N$ is increased. This shows that the ground energy has been raised after the correction of the nuclear motion is considered. One can see from the data in Table 1 and Table 2 that it is necessary to consider the relativistic and radiative corrections to the nonrelativistic energy.

Finally, considering all these contributions, we get the high precision value of the helium ground energy $-2.90338$ a.u. and the relative error is $0.00034{\%}$ (the experimental value is $-2.90338629$ a.u. \cite{Radzing1985}).

In these calculations, we use a Mathematica software, which is a tool of symbolic calculation. All results are exact except for that of the last step that will bring an error. But this error can be controlled by setting the precision, and we have ensured that it is much smaller than $\Delta E$. So, the error is mainly derived from the inaccurateness of model, i.e. the ignorer of the higher orders relativistic and QED effects. We take the error as half of order $\alpha ^{3}$ relativistic correction and the estimated theoretical uncertainty of calculation is obtained to be $\pm 0.00001 $ a.u..

In conclusion, we pointed out that the result of variational calculation of the ground state energy for a helium atom would lower than the experimental value when the nuclear motion is ignored. First, a new variational wave function with a flexible scaling parameter $k $has been constructed. Then we take account of the nuclear motion and the relativistic and radiative corrections to the nonrelativistic ground energy. Using the variational method and perturbation method, the high precision value of the helium ground energy is worked out. It is obvious that accuracy will be improved increasing with the number of basis functions $N$. Thus the ground energy problem of a helium atom is now solved satisfactorily. In our later work, higher orders relativistic and QED effects will be considered to agree with the observational value better.

The authors are grateful to Yu-Hong Chen, Yong-Qiang Wang and Zhen-Hua Zhao for their useful discussions.

\end{document}